\documentclass{SCIS2023}

\begin{document}
\ArticleType{RESEARCH PAPER}
\Year{2022}
\Month{}
\Vol{}
\No{}
\DOI{}
\ArtNo{}
\ReceiveDate{}
\ReviseDate{}
\AcceptDate{}
\OnlineDate{}

\title{Practical Continuous-variable Quantum Key Distribution with Feasible Optimization Parameters}{Practical Continuous-variable Quantum Key Distribution with Feasible Optimization Parameters}

\author[1]{Li Ma}{}
\author[1,2]{Jie Yang}{}
\author[1]{Tao Zhang}{}
\author[1]{Yun Shao}{}
\author[1]{Jinlu Liu}{}
\author[1]{Yujie Luo}{}
\author[1]{Heng Wang}{}
\author[1]{Wei Huang}{}
\author[1]{\\Fan Fan}{}
\author[1]{Chuang Zhou}{}
\author[1]{Liangliang Zhang}{}
\author[1]{Shuai Zhang}{}
\author[2]{Yichen Zhang}{}
\author[1]{Yang Li}{{yishuihanly@pku.edu.cn}}
\author[1]{Bingjie Xu}{{xbjpku@pku.edu.cn}}

\AuthorMark{Li Ma}

\AuthorCitation{Li Ma, Jie Yang, Tao Zhang, et al}


\address[1]{Science and Technology on Communication Security Laboratory,\\ Institute of Southwestern Communication, Chengdu {\rm 610041}, China}
\address[2]{State Key Laboratory of Information Photonics and Optical Communications,\\Beijing University of Posts and Telecommunications, Beijing {\rm 100876}, China}

\abstract{Continuous-variable quantum key distribution (CV-QKD) offers an approach to achieve a potential high secret key rate (SKR) in metropolitan areas. There are several challenges in developing a practical CV-QKD system from the laboratory to the real world. One of the most significant points is that it is really hard to adapt different practical optical fiber conditions for CV-QKD systems with unified hardware. Thus, how to improve the performance of practical CV-QKD systems in the field without modification of the hardware is very important. Here, a systematic optimization method, combining the modulation variance and error correction matrix optimization, is proposed to improve the performance of a practical CV-QKD system with a restricted capacity of postprocessing. The effect of restricted postprocessing capacity on the SKR is modeled as a nonlinear programming problem with modulation variance as an optimization parameter, and the selection of an optimal error correction matrix is studied under the same scheme. The results show that the SKR of a CV-QKD system can be improved by 24\% and 200\% compared with previous frequently used optimization methods theoretically with a transmission distance of 50 $km$. Furthermore, the experimental results verify the feasibility and robustness of the proposed method, where the achieved optimal SKR achieved practically deviates $<$1.6\% from the theoretical optimal value. Our results pave the way to deploy high-performance CV-QKD in the real world.}


\maketitle

\section{Introduction}
Quantum key distribution (QKD) can realize secure key distribution remotely with unsecured channels in real time based on the principle of quantum mechanics, which has made a series of progresses in recent years \cite{r1, r2,ra10,ra11,ra12,ra13,ra14,ra15,ra16}. There are mainly two types of QKD protocols, which respectively encode information on discrete variables \cite{r3,r4} and continuous variables \cite{r5,r6,ra9}. The continuous-variables QKD (CV-QKD) takes advantage of the use of standard telecommunication technologies \cite{r7,r8,r9,r10} and obtains high key rates within metropolitan areas. 

Recently, significant progress has also been made in the field of theory and experiments on CV-QKD. On the one hand, each core procedure of a CV-QKD system (e.g., quantum state preparation, measurement, and postprocessing) should be correctly and efficiently implemented experimentally, which provides a \textbf{hardware basis} for high-performance CV-QKD and has made significant progress in recent years \cite{r11,r12,r13,ra17,ra18,r14,r15,r16,r17,r18,r19}. On the other hand, to further enhance the performance of a CV-QKD system, many advanced theoretical methods have been proposed and demonstrated, including the excess noise modeling and suppression \cite{r8,r20,r21,r22}, system parameter optimization method (e.g., modulation variance $V_A$), advanced information reconciliation method \cite{r23,r24,r25,r26,r27,ra1}, high-efficiency error correction matrix design \cite{r28,r29}, rate-adaptive algorithm \cite{r30,ra2,ra3,ra4,ra5,ra6}, postselection \cite{r31,r32,r33}, and add noise method \cite{r34}, which provides a \textbf{software basis} to improve the system performance. Based on a particular CV-QKD hardware setup with a predefined protocol, how to comprehensively optimize the secret key rate (SKR) based on the above theoretical methods are of great importance. However, the above methods are not independent of one another, and sometimes one needs to make a systematic optimization. For example, the optimal choice of $V_A$ will significantly influence the SKR, which is closely related to the error correction matrix \textbf{H} design and choice. A global optimization method for a CV-QKD system in software is still incomplete, among which the systematic optimization of system parameters with error correction is of special importance with respect to the SKR. In contrast, there are several challenges in developing widely used practical CV-QKD systems. One of the most significant points is that it is really hard to adapt different practical optical fiber conditions for CV-QKD systems with unified hardware. Thus, how to improve the performance of practical CV-QKD systems in the field without modification of the hardware is very important.

In a practical CV-QKD system, $V_A$ should be adjusted periodically with the variation in system parameters to optimize the system performance in real time. When the system environment is relatively stable in the short term, the system parameters channel transmittance $T$, excess noise $\xi$, detection electrical noise $v_{el}$, and detection efficiency $\eta$ change slowly with time. In this condition, one can reasonably use the calculated optimal $V_A$ based on the measured system parameters of a raw data block as the expected real optimal $V_A$ for the next successive raw data block. Thus, practically, the signal-to-noise ratio (SNR) of the system is mainly determined by $V_A$, which directly decides the performance of the data reconciliation in postprocessing and significantly affects the SKR of the CV-QKD system. Although various schemes to optimize  $V_A$ have been proposed, it is usually assumed that the reconciliation efficiency $\beta$ in data reconciliation and frame error rate (FER) in error correction are constant under a specific transmission distance \cite{r16,r24}. However, since the SNR is mainly determined by $V_A$ under a specific transmission distance, to maintain $\beta$ as a constant for different $V_A$, several different \textbf{H} with corresponding code rates should be designed for the data reconciliation, which is very difficult to fulfill in practice. Furthermore, the FER for a specific \textbf{H} should vary under different $V_A$ \cite{r35}, and various \textbf{H} are needed to keep the FER as a constant, which is difficult to fulfill. Actually, under typical transmission distances, it is only possible to switch between a few well-designed \textbf{H} with different code rates according to the SNRs \cite{r8}. The fluctuations of $\beta$ and FER with $V_A$ should be taken into consideration to achieve a systematic optimization.

In this paper, we propose and experimentally verify a feasible optimization method for a practical CV-QKD system with a restricted capacity of postprocessing. Different from the previously proposed method, in our work, for the certain data reconciliation and error correction matrix \textbf{H}, the influences of $V_A$ on $\beta$ and the FER are quantitatively analyzed and experimentally verified. This method can be easily applied in a practical CV-QKD system, and the actual achieved SKR deviates only $<$1.6\% from the theoretical optimal value in our experiment. Moreover, the proposed method can be combined with various advanced postprocessing technologies to realize a global optimization method for the CV-QKD system in software without any hardware modification, such as rate-adaptive algorithm, postselection, and add noise method, to further improve the performance of CV-QKD. 

The paper is organized as follows: In Section ~\ref{sec:level2}, we introduce the optimization method for the CV-QKD system with GG02 and no-switching protocol, respectively. In Section ~\ref{sec:level3}, we show the simulation and experimental performance of the method. Finally, in Section ~\ref{sec:level4},we discuss and conclude this paper.

\section{\label{sec:level2}Feasible optimization method for a CV-QKD system}
First, the influence of $V_A$ on the SKR for GG02 \cite{r5} and no-switching CV-QKD protocol \cite{r6} is quantitatively analyzed, where the optimization of the SKR can be defined as a nonlinear programming problem. Second, an experimentally feasible operational process for the above optimization method is given, where the fitting curve of the FER on $V_A$ is experimentally given.

\subsection{\label{sub:level1}Theoretical model for the optimal choice of $V_A$ and the error correction matrix}
In the GG02 protocol with homodyne detection, we suppose that $x$ and $y$ denote the raw keys of Alice and Bob after the sifting process. The SKR with composable finite-size security can be expressed as \cite{r36,ra7}
\begin{equation}\label{eq1}
\begin{aligned}
K^{hom}_{finite}=\frac{np_{ec}}{N}(\beta^{hom}I^{hom}(x:y)-\chi^{hom}(y:E)-\frac{\Delta_{aep}}{\sqrt{n}}+\frac{\Theta}{\sqrt{n}})
\end{aligned}
\end{equation}
where $N$ is the block size of raw key, $n$ ($m=N-n$) is the fraction for key distillation (parameter estimation), and the probability of successful error-correction is $p_{ec}$ ($p_{ec}=1-FER$), $FER\in[0,1]$, $\beta^{hom}\in[0,1]$, $I^{hom}(x:y)$ is the Shannon mutual information between Alice and Bob, $\chi^{hom}(y:E)$ is the Holevo bound of Bob and Eve. The extra finite-size terms are given as

\begin{equation}\label{req1}
\begin{aligned}
\Delta_{aep}=4log_2(\sqrt{d}+2)\sqrt{log_2\left(\frac{18}{p_{ec}^2\varepsilon_s^4}\right)}
\end{aligned}
\end{equation}

\begin{equation}\label{req2}
\begin{aligned}
\Theta=log_2\left[p_{ec}\left(1-\frac{\varepsilon_s^2}{3}\right)\right]+2log_2\sqrt{2}\varepsilon_h
\end{aligned}
\end{equation}
where $d$ representing the size of the effective alphabet after analog-to-digital conversion of sender's and receiver's CVs (quadrature encodings and outcomes). $\varepsilon_h$ is a hashing parameter and $\varepsilon_s$ is a smoothing parameter with a value of $10^{-10}$. Reverse reconciliation is employed here to beat the 3dB limit in CV-QKD.

As shown in ~\ref{app:subsec}, $I^{hom}(x:y)$ and $\chi^{hom}(y:E)$ can be analytically represented by $T$, $\xi$, $\eta$, $v_{el}$, and $V_A$. When the system environment is relatively stable, $T$, $\xi$, $\eta$, and $v_{el}$ should change slowly which can be approximated as constants in two successive rounds of raw data. Thus $I^{hom}(x:y)$ and $\chi^{hom}(y:E)$ can be treated as a function of the $V_A$, where $I^{hom}(x:y)=f_{I^{hom}(x:y)}(V_A)$ and $\chi^{hom}(y:E)=f_{\chi^{hom}(y:E)}(V_A)$. $\beta^{hom}=R/I^{hom}(x:y)$, where $R$ is the code rate of the practically used error correction matrix \textbf{H}. Furthermore, the performance of the decoding in error correction is directly related to the $SNR$, which means $FER=f_{FER}(V_A)$ is decided by \textbf{H} and can be measured experimentally. Denote $Q=-\frac{\Delta_{aep}}{\sqrt{n}}+\frac{\Theta}{\sqrt{n}}$ in Eqs.~(\ref{eq1}) -~(\ref{req2}), and one can easily verify that $Q$ is directly related to $FER$ given fixed $n$, $d$, $\varepsilon_h$, and $\varepsilon_s$, which means $Q=f_Q(V_A)$.

As a result, the SKR can be expressed as

\begin{equation}\label{eq2}
\begin{aligned}
K^{hom}_{finite}(V_A)=f(V_A)^{hom}=\frac{n}{N}(1-f_{FER}(V_A))(R-f_{\chi^{hom}(y:E)}(V_A)+f_Q(V_A))
\end{aligned}
\end{equation}

In the no-switching protocol with heterodyne detection, the SKR with composable finite-size security can be expressed as \cite{r36,ra7}

\begin{equation}\label{eq3}
\begin{aligned}
K^{het}_{finite}=\frac{np_{ec}}{N}(\beta^{het}I^{het}(x:y)-\chi^{het}(y:E)-\frac{\Delta_{aep}}{\sqrt{n}}+\frac{\Theta}{\sqrt{n}})
\end{aligned}
\end{equation}
Similarly, $I^{het}(x:y)=log_2(1+SNR)$, $\beta^{het}I^{het}(x:y)=2R$, and $\chi^{het}(y:E)=f_{\chi^{het}(y:E)}(V_A)$ .

Thus, the SKR for no-switching protocol can be expressed as
\begin{equation}\label{eq4}
\begin{aligned}
K^{het}_{finite}(V_A)=f(V_A)^{het}=\frac{n}{N}(1-f_{FER}(V_A))(2R-f_{\chi^{het}(y:E)}(V_A)+f_Q(V_A))
\end{aligned}
\end{equation}
In both Eq.~(\ref{eq2}) and Eq.~(\ref{eq4}), the $V_A$ can be regarded as the only variable when $T$, $\xi$, $\eta$, and $v_{el}$ remain stable in two successive raw data block. Therefore, the optimization of the SKR for a given \textbf{H} is a constrained nonlinear programming problem as

\begin{equation}\label{eq5}
\underset{V_A}{Max}\frac{n}{N}(1-f_{FER}(V_A))(v_{det}R-f_{\chi^{hom/het}(y:E)}(V_A)+f_Q(V_A))
\end{equation}
where $v_{det}$ is the quantum duty ("qu-duty") associated with detection: $v_{det}=1$ for homodyne and $v_{det}=2$ for heterodyne. Subject to $0\le\beta^{hom/het}\le1$, where $f_{FER}(V_A)$, $f_{\chi^{hom/het}(y:E)}(V_A)$, and $f_Q(V_A)$ are nonlinear functions of $V_A$. For a given  \textbf{H}, the optimization method therefore realizes a trade-off between the frame error rate $f_{FER}(V_A)$ and the Holevo information $f_{\chi^{hom/het}(y:E)}(V_A)$.

Based on the above method, one can easily make a systematic optimal choice of $V_A$ and \textbf{H} with different code rates under different transmission distances.

\subsection{\label{sub:level2}Feasible operational process for the proposed optimization method}
\begin{figure}
\centering  
\includegraphics[width = 16cm]{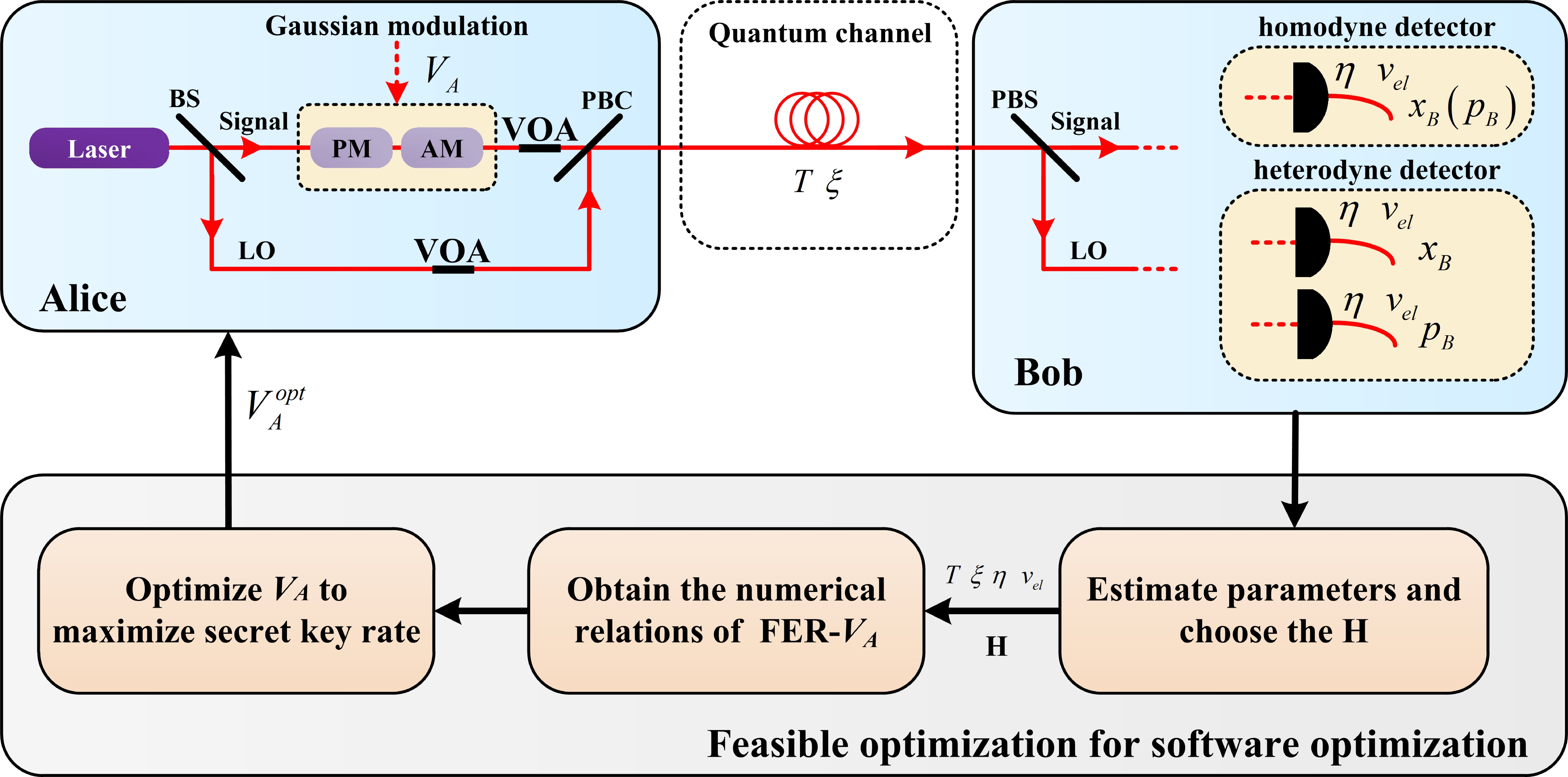}
\caption{\label{fig1}Feasible optimization scheme for the proposed optimization method. For a heterodyne detector, Bob employs two identical balanced detectors for simplicity. BS: beam splitter, AM: amplitude modulator, PM: phase modulator, PBS/C: polarization beam splitter/coupler, LO: local oscillator, VOA: variable optical attenuator.}
\end{figure}
The feasible optimization scheme for the proposed method in a CV-QKD system is shown in Fig.~\ref{fig1}. First, we obtained the system parameters $\{T, \xi, \eta, v_{el}\}$ and chose an appropriate \textbf{H}. Usually, we choose the range of $V_A$ from 0 to 100 \cite{r38}, and we expect $\beta$ to be as large as possible, e.g., $\beta\in[0.8,1]$. Thus, the SNR range can be roughly estimated, and the corresponding \textbf{H} can be preliminarily chosen. For a CV-QKD system with a relatively stable environment, $T, \xi, \eta, v_{el}$ should change slowly in a period of time, which can be measured and updated in real time for each data block. Second, we obtained the numerical relation of $FER-V_A$ via curve fitting for the chosen \textbf{H}. Based on the specific performance of data reconciliation and error correction, $f_{FER}(V_A)$ can be obtained experimentally. Finally, we substituted $f_{FER}(V_A)$ into Eqs.~(\ref{eq2}) and ~(\ref{eq4}) to obtain the comprehensive function of SKR on $V_A$, based on which the optimal modulation variance $V_A^{opt}$ can be accordingly estimated.

In the following, we introduce in detail the method to obtain the numerical relation of $f_{FER}(V_A)$. In homodyne detection, the same system parameters as in Ref. \cite{r24,r39} are employed to numerically generate raw data. The transmission distance was set as $L$=50 $km$, $\eta$=0.606, $\xi$=0.005, $v_{el}$=0.041, $T=10^{-\alpha L/10}$ with $\alpha$=0.2 $dB/km$. The degree distribution function with $R$=0.1 as in Ref. \cite{r30} was chosen to generate the \textbf{H} using our own matrix generation method as in Ref. \cite{r40}. $f_{FER}(V_A)$ was directly determined by the SNR and \textbf{H}. Thus, with different \textbf{H}, different results will be obtained.

\begin{figure}
    {\label{fig2a}{\includegraphics[width = 8cm,height = 4.3cm]{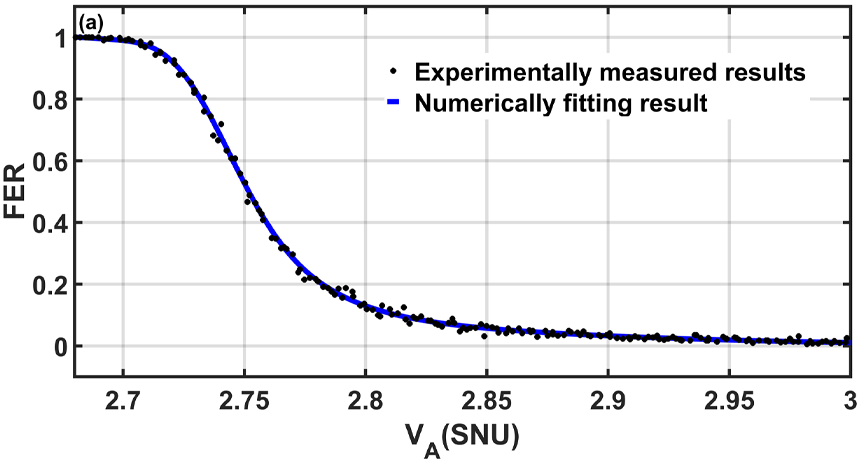}}}
    {\label{fig2b}{\includegraphics[width = 8cm,height = 4.3cm]{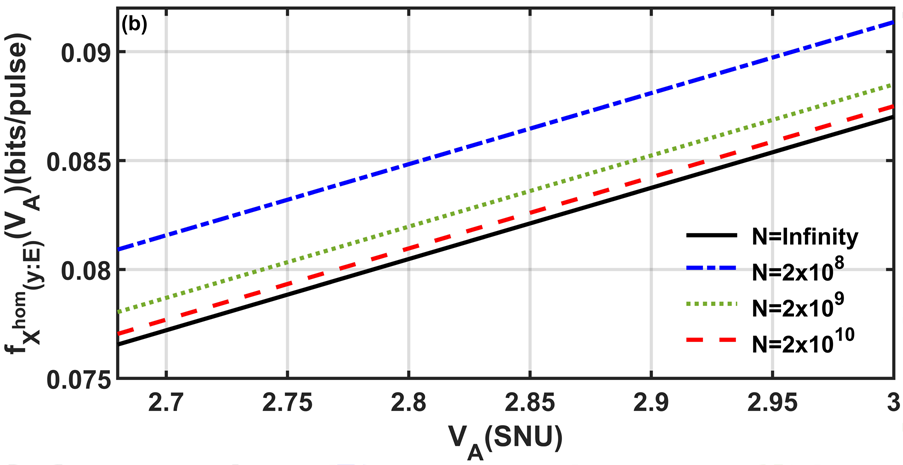}}}
    \caption{\label{fig2}(a) Experimentally measured results (black dots) and numerical fitting result (blue line) of $FER-V_A$ with the GG02 protocol. (b) Simulation curves of the $f_{\chi^{hom}(y:E)}(V_A)$ with the GG02 protocol. From top to bottom, the curves respectively show the results with a total block size of $N=2\times10^8$, $2\times10^9$, $2\times10^{10}$, and $\infty$. The results show that both of them are proportional to $V_A$. The system parameters are set as follows: $L=50$ $km$, $\eta=0.606$, $\xi$=0.005, $v_{el}$=0.041, $\alpha$=0.2 $dB/km$, $T$=0.1, and $R$=0.1. SNU: shot noise units. The number of iterations is 60.}
\end{figure}

To measure the function $FER-V_A$, we first generated raw data based on the above system parameters with $V_A\in(0,100]$. Second, an eight-dimensional multidimensional reconciliation was performed on the raw data, and the chosen \textbf{H} was employed for error correction. Thus, the FERs in the error correction under different $V_A$ can be measured experimentally, and the numerical relationship $FER-V_A$ can be obtained for a practical CV-QKD system. The curve fitting result of $FER-V_A$ to generate \textbf{H} is shown in Fig.~\ref{fig2}(a). When $0<V_A<2.7$, $FER$=1, indicating that the error correction has all failed. When $V_A>3$, $FER$=0, indicating that the error correction has all succeeded. However, when $2.7\le V_A\le3$, $FER$ decreases with the increase in $V_A$. Therefore, we mainly needed to perform the curve fitting in the range of $2.7\le V_A\le3$. By employing different fitting functions and comparing the corresponding fitting effects and computational complexities, finally, the fourth-order Gaussian function was chosen in this paper, as shown in Eq.~(\ref{eq6})
\begin{equation}\label{eq6}
\begin{aligned}
&FER=f_{FER}(V_A)\\
&=\left\{\begin{matrix}
 1,       &0<V_A<2.7 \\\\ 
0.8310e^{-(\frac{V_A-2.654}{0.08704})^2}+0.6753e^{-(\frac{V_A-2.113}{0.4542})^2}+0+0.3437e^{-(\frac{V_A-2.722}{0.03649})^2},&2.7\le V_A\le3\\\\
0,        &V_A>3 
\end{matrix}\right.
\end{aligned}
\end{equation}

\begin{figure}
    {\label{fig3a}{\includegraphics[width = 8cm,height = 4.3cm]{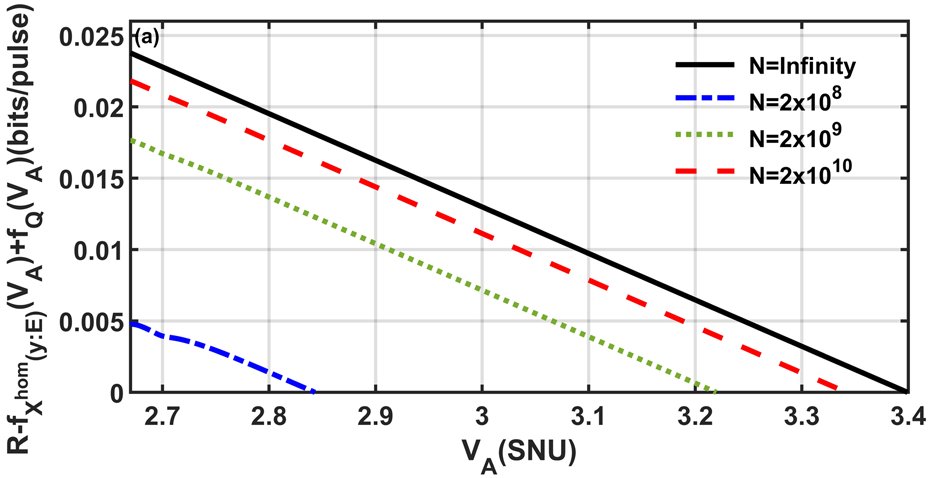}}}
    {\label{fig3b}{\includegraphics[width = 8cm,height = 4.3cm]{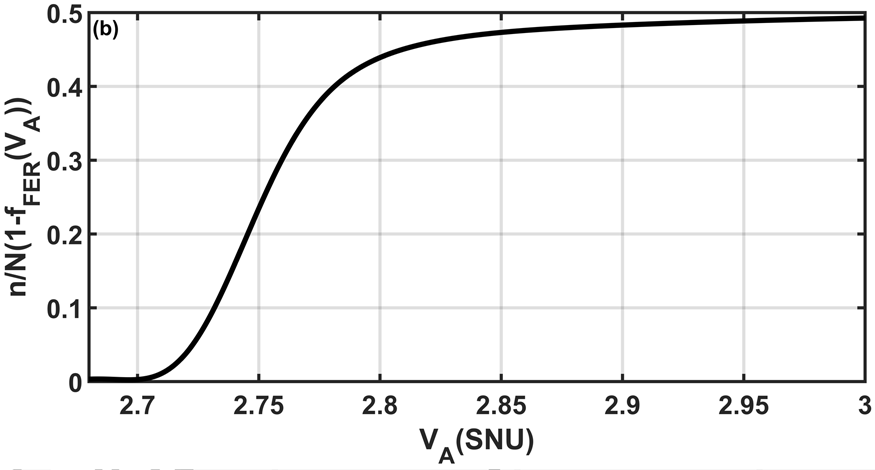}}}
	 \caption{\label{fig3}(a) Simulation curves of the $R-f_{\chi^{hom}(y:E)}(V_A)+f_Q(V_A)$ with respect to $V_A$ under the GG02 protocol. From bottom to top, the curves respectively show the results with the total block size of $N=2\times10^8$, $2\times10^9$, $2\times10^{10}$, and $\infty$. The results show that both of them are inversely proportional to $V_A$. (b) The curve is $\frac{n}{N}(1-f_{FER}(V_A))$ with respect to $V_A$. The system parameters are set as follows: $\frac{n}{N}=0.5$, $L=50$ $km$, $\eta=0.606$, $\xi=0.005$, $v_{el}=0.041$, $\alpha=0.2$ $dB/km$, $T=0.1$, and $R=0.1$. SNU: shot noise units.}
\end{figure}

In the numerical simulation, the block size $N$ is set as $2\times10^8$, $2\times10^9$, $2\times10^{10}$, and $\infty$, respectively, and $n$ is set to be $N/2$. The curve of $f_{\chi^{hom}(y:E)}(V_A)$ is shown in Fig.~\ref{fig2}(b), which is a monotonically increasing function of $V_A$. Furthermore, the simulation curves of $R-f_{\chi^{hom}(y:E)}(V_A)+f_Q(V_A)$ and $\frac{n}{N}(1-f_{FER}(V_A))$ with respect to $V_A$ for GG02 protocol is shown in Fig.~\ref{fig3}, where $R-f_{\chi^{hom}(y:E)}(V_A)+f_Q(V_A)$ is a monotonically decreasing function of $V_A$, and $\frac{n}{N}(1-f_{FER}(V_A))$ is a monotonically increasing function of $V_A$. Therefore, the optimal value of $V_A$ to obtain the maximum SKR can be found.

Finally, we substitute Eq.~(\ref{eq6}) into Eq.~(\ref{eq2}) to obtain $K^{hom}_{finite}(V_A)=f(V_A)^{hom}$.

\begin{figure}
    {\label{figr4a}{\includegraphics[width = 8cm,height = 4.3cm]{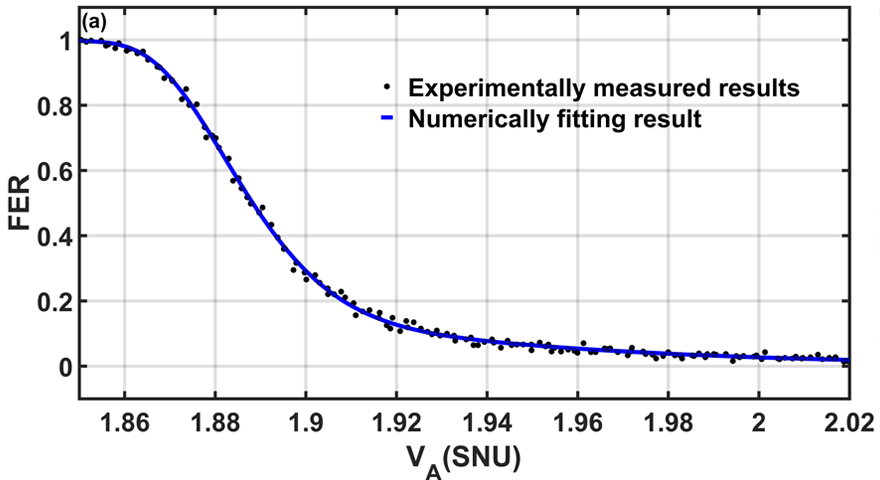}}}
    {\label{figr4b}{\includegraphics[width = 8cm,height = 4.3cm]{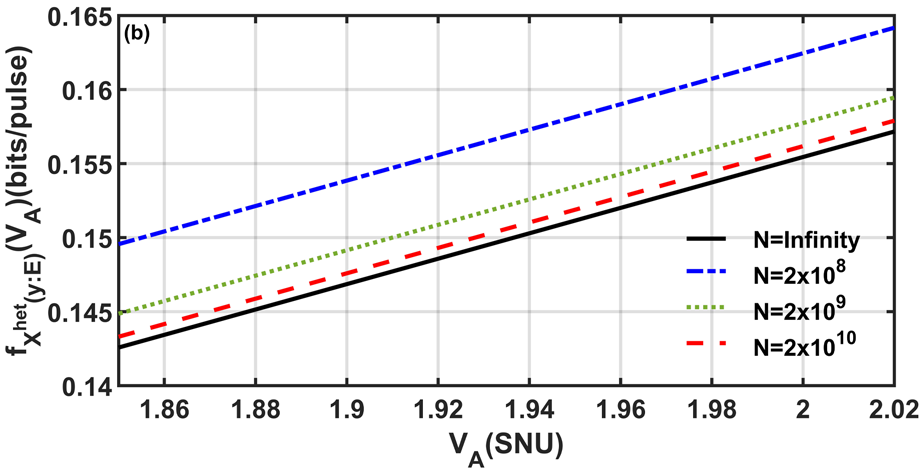}}}
\caption{\label{figr4}(a) Experimental measurement results (black dots) and numerical fitting results (blue line) of $FER-V_A$ with the no-switching protocol. (b) Simulation curves of $f_{\chi^{het}(y:E)}(V_A)$ with the no-switching protocol. From top to bottom, the curves respectively show the results with the total block size of $N=2\times10^8$, $2\times10^9$, $2\times10^{10}$, and $\infty$. The results show that both of them are proportional to $V_A$. The system parameters are set as follows: $L$=25 $km$, $\eta$=0.56, $\xi$=0.022, $v_{el}$=0.042, $\alpha$=0.2 $dB/km$, $T$=0.3162, and $R$=0.1. SNU: shot noise units. The number of iterations is 60.}
\end{figure}

\begin{figure}
    {\label{figr5a}{\includegraphics[width = 8cm,height = 4.3cm]{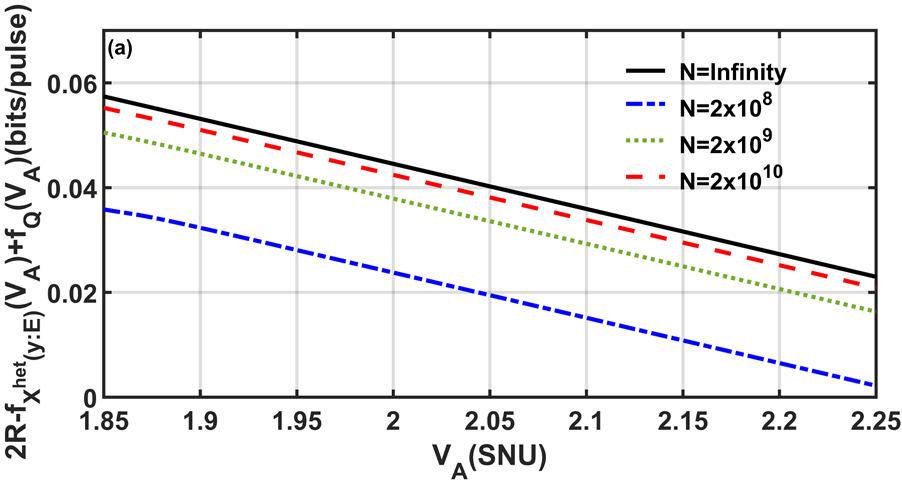}}}
    {\label{figr5b}{\includegraphics[width = 8cm,height = 4.3cm]{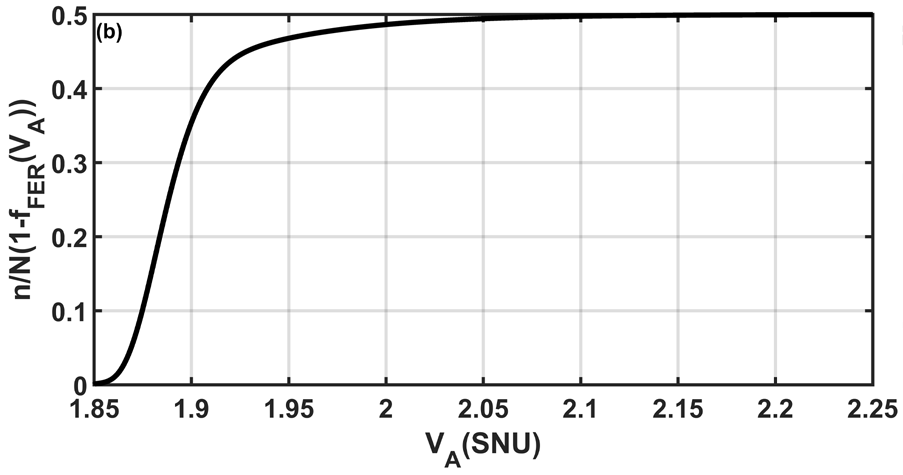}}}
\caption{\label{figr5}(a) Simulation curves of $2R-f_{\chi^{het}(y:E)}(V_A)+f_Q(V_A)$ with respect to $V_A$ under the no-switching protocol. From bottom to top, the curves respectively show the results with the total block size of $N=2\times10^8$, $2\times10^9$, $2\times10^{10}$, and $\infty$. The results show that both of them are inversely proportional to $V_A$. (b) The curve is $\frac{n}{N}(1-f_{FER}(V_A))$ with respect to $V_A$. The system parameters are set as follows: $\frac{n}{N}=0.5$, $L=25$ $km$, $\eta=0.56$, $\xi=0.022$, $v_{el}=0.042$, $\alpha=0.2$ $dB/km$, $T=0.3162$, and $R=0.1$. SNU: shot noise units.}
\end{figure}

Likewise, for CV-QKD system with heterodyne detection, the system parameters \cite{r13} are set as $L$=25 $km$, $\eta$=0.56, $\xi$=0.022, $v_{el}$=0.042, $\alpha$=0.2 $dB/km$, $T$=0.3162, and $R$=0.1. The numerical relation of $FER-V_A$ can be obtained in same way. The curve fitting result of $FER-V_A$ obtained with the fourth-order Gaussian function is shown in Eq.~(\ref{eq7}) and the corresponding fitting curve is shown in Fig.~\ref{figr4}(a). The curve of $f_{\chi^{het}(y:E)}(V_A)$, $2R-f_{\chi^{het}(y:E)}(V_A)+f_Q(V_A)$, and $\frac{n}{N}(1-f_{FER}(V_A))$ for no-switching protocol are shown in Fig.~\ref{figr4}(b), Fig.~\ref{figr5}.

\begin{equation}\label{eq7}
\begin{aligned}
&FER=f_{FER}(V_A)\\
=&\left\{\begin{matrix}
 1,       &0<V_A<1.84 \\\\ 
-0.1987e^{-(\frac{V_A-1.851}{0.01752})^2}-0.8834e^{-(\frac{V_A-1.854}{0.03432})^2}+0+8727e^{-(\frac{V_A-0.5462}{0.4082})^2},&1.84\le V_A\le2.02\\\\
0,        &V_A>2.02 
\end{matrix}\right.
\end{aligned}
\end{equation}

Finally, we substitute Eq.~(\ref{eq7}) into Eq.~(\ref{eq4}) to obtain $K^{het}_{finite}(V_A)=f(V_A)^{het}$.

\subsection{\label{sub:level3}Feasibility of the method}
To calculate the optimal $V_A$, two kinds of information are needed: system parameters ($T$, $\xi$, $\eta$, and $v_{el}$) and $f_{FER}(V_A)$. The system parameters can be efficiently estimated in real time in the parameter estimation process. Although $f_{FER}(V_A)$ is fitted numerically without analytic solutions, $f_{FER}(SNR)$ is \textbf{only} determined by the error correction matrix \textbf{H} and decoding method. After a specifically optimal \textbf{H} and the decoding method is chosen for a CV-QKD system under a certain transmission distance, $f_{FER}(SNR)$ is only needs to be fitted numerically once, where collecting fitting curves with respect to all possible parameters is unnecessary. In a real CV-QKD experiment, $f_{FER}(V_A)$ can be easily calculated in real time through the function $f_{FER}(SNR)$ and parameters ($T$, $\xi$, $\eta$, $v_{el}$), where $SNR=\frac{V_A\eta T}{\eta T\xi+v_{det}+v_{det}v_{el}}$. Therefore, the proposed method can deal with the time-varying parameters, which can be effectively applied in a practical CV-QKD system.

\section{\label{sec:level3}Performance of the proposed optimization method}
In the following, the performance of the optimization method is verified through a numerical simulation and experimental test, where a CV-QKD system with the GG02 protocol is implemented to present the optimization results experimentally.
\subsection{\label{sub:level3}Numerical simulation}

\begin{figure}
\centerline{
\includegraphics[width = 15cm]{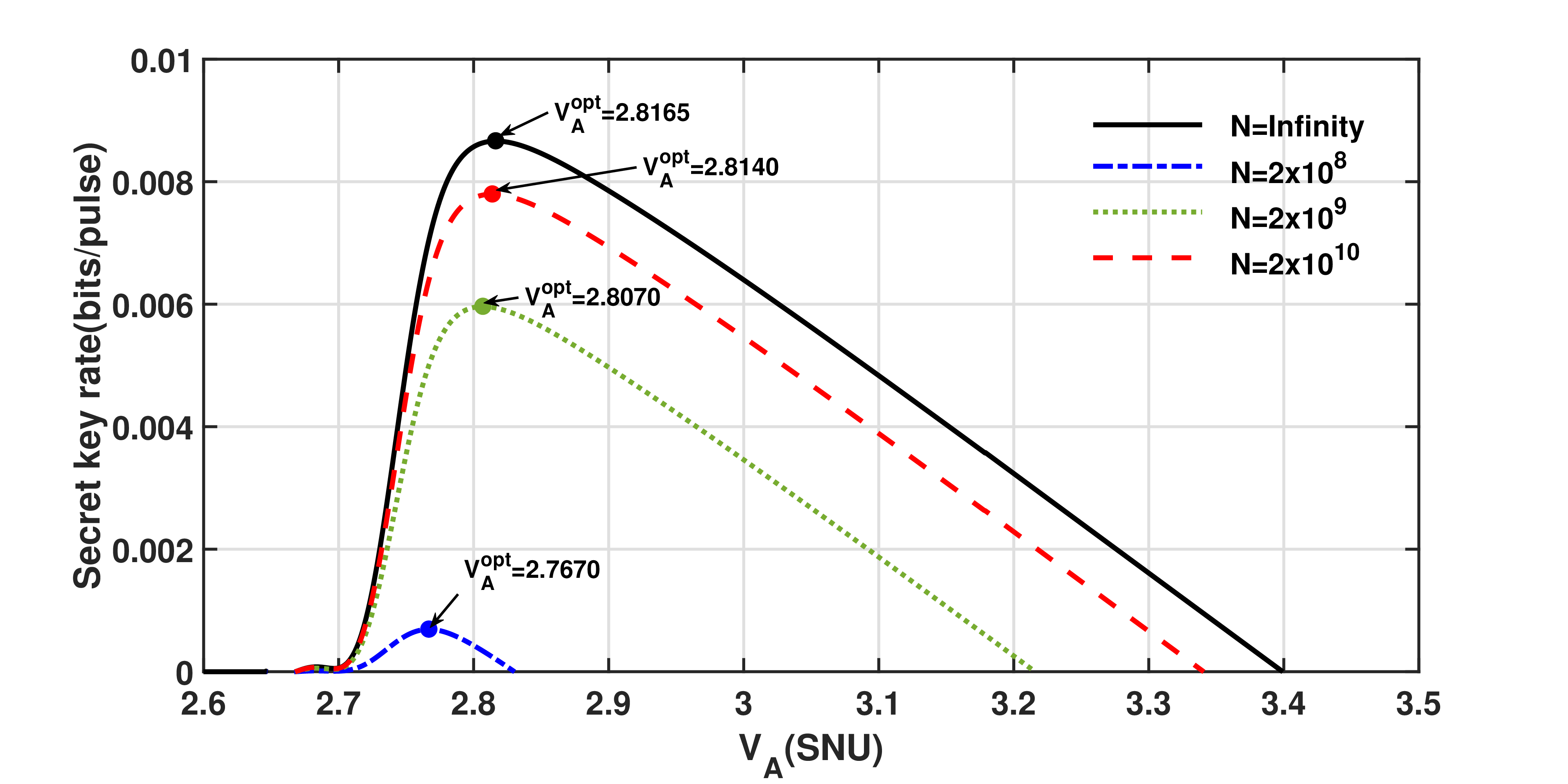}}

\caption{\label{fig4}Simulation curves of the SKR varying with $V_A$ under the GG02 protocol. From bottom to top, the curves respectively show the results with the total block size of $N=2\times10^8$, $2\times10^9$, $2\times10^{10}$, and $\infty$. $V_A^{opt}$ are 2.7670, 2.8070, 2.8140, and 2.8165. The system parameters are set as follows: $L$=50 $km$, $\eta$=0.606, $\xi$=0.005, $v_{el}$=0.041, $\alpha$=0.2 $dB/km$, $T$=0.1, and $R$=0.1. SNU: shot noise units.}
\end{figure}
\begin{figure}
\centerline{
\includegraphics[width = 15cm]{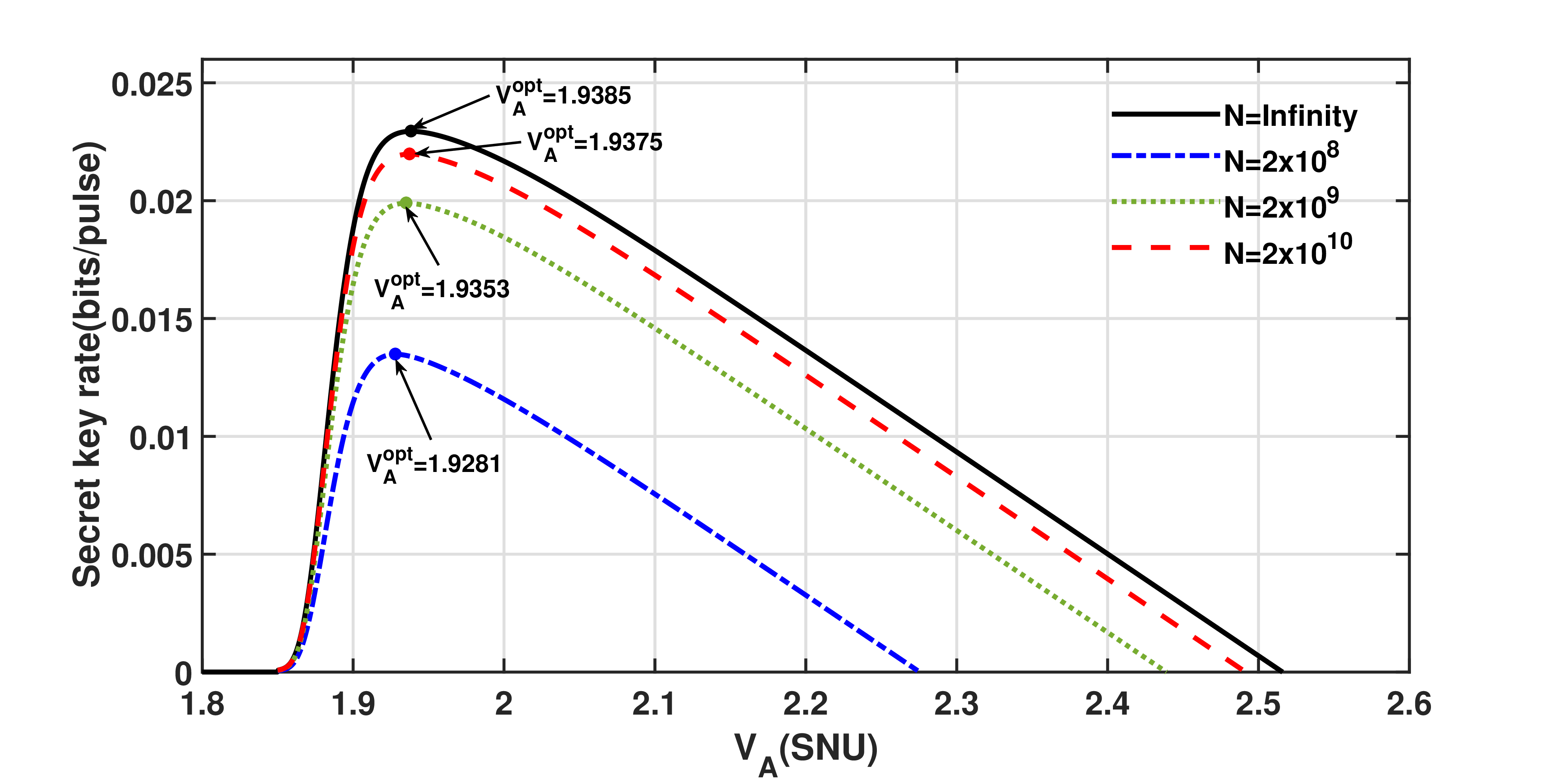}}

\caption{\label{fig5} Simulation curves of the SKR varying with $V_A$ under the no-switching protocol. From bottom to top, the curves respectively show the results with the total block size of $N=2\times10^8$, $2\times10^9$, $2\times10^{10}$, and $\infty$. $V_A^{opt}$ are 1.9281, 1.9353, 1.9375, and 1.9385. The system parameters are set as follows: $L$=25 $km$, $\eta$=0.56, $\xi$=0.022, $v_{el}$=0.042, $\alpha$=0.2 $dB/km$, $T$=0.3162, and $R$=0.1. SNU: shot noise units.}
\end{figure}

In the numerical simulation, the block size $N$ was set as $2\times10^8$, $2\times10^9$, $2\times10^{10}$, and $\infty$, and $n$ was set to $N/2$. Then, by calculating the SKR under different $V_A$, the optimal $V_A^{opt}$ under homodyne detection can be estimated, as shown in Fig.~\ref{fig4}. For the block sizes of $2\times10^8$, $2\times10^9$, $2\times10^{10}$, and $\infty$, $V_A^{opt}$ was set as 2.7670, 2.8070, 2.8140, and 2.8165, and the corresponding optimal SKRs were 0.0007 $bits/pulse$, 0.0060 $bits/pulse$, 0.0078 $bits/pulse$, and 0.0087 $bits/pulse$, respectively. The results show that $V_A^{opt}$ gradually enhanced with the increase in $N$, but the change is almost negligible. Similarly, the simulation results under heterodyne detection are shown in Fig.~\ref{fig5}.

Taking $N=\infty$ as an example, the performance of our optimization method was compared with that of two frequently used optimization methods presented in Ref. \cite{r23,r24} and Ref. \cite{r16,r38,r39}. In the first method, to adapt the code rates of various error correction matrices \textbf{H} under different transmission distances, $V_A$ was accordingly adjusted to guarantee several fixed SNRs, which is referred to as the first method. In the second method, $\beta$ and FER are assumed to be constant, and then $V_A$ was optimized by maximizing SKR, which is referred to as the second method. In the numerical simulation, for comparison, the same data reconciliation and error correction method were employed, and the simulation parameters were set as $L$=$50$ $km$, $\eta$=0.606, $\xi$=0.005, $v_{el}$=0.041, $\alpha$=0.2 $dB/km$, $T$=0.1, and $R$=0.1.
The simulation results are shown in Table~\ref{tab:tab1}. For the first method, the SNR was fixed as 0.161 with $\beta$=92.85\% and $V_A$=2.7665. The experimentally verified result of the FER with our \textbf{H} in this case is 0.3192, and the SKR is 0.0070 $bits/pulse$. For the second method, $\beta_0$ and $FER_0$ were assumed to be 92\% and 0.1, respectively. However, the optimization for $V_A$ was performed without comprehensively considering the performance of the postprocessing. Consequently, the expected theoretical values of $\beta$ and FER cannot be practically achieved. $V_A^{opt}$ was 3.2193, the corresponding SNR was 0.1874, and the experimentally verified $\beta$ was 80.71\%, resulting in an SKR of 0.0029 $bits/pulse$. Finally, for the method proposed in this work, the SKR was 0.0087 $bits/pulse$ while comprehensively considering the influence of $V_A$ on the FER and $\beta$. Accordingly, an improvement of 24.29\% and 200\% compared with the first and second methods, respectively, was achieved. By calculating the PLOB bound \cite{ra8}, the SKR optimization result of our method achieved an improvement of 0.9443 $dB$ and 4.7713 $dB$ compared with the first and second methods, respectively.

\begin{table}
\tabcolsep 5pt 
\caption{\label{tab:tab1}Comparison of SKR optimization results with different methods. For method one, the SNR is fixed in 0.161 with $\beta$=92.85\%. For method two, the $\beta_0$ and $FER_0$ are expected value of  $\beta$ and $FER$, which are assumed to be 92\% and 0.1, respectively. $R$: code rate; SNR: signal noise ratio; FER: frame error rate. The system parameters are set as: $L$=50$km$, $\eta$=0.606, $\xi$=0.005, $v_{el}$=0.041, $\alpha$=0.2$dB/km$, $T$=0.1 and $R$=0.1.}
\begin{tabular}{cccccccc}
\toprule
 & \bm{$R$} & \bm{$V_A$} &\bm{$\beta(\%)$} & \bm{$SNR$}& \textbf{FER}& \textbf{SKR(bit/pulse)}& \textbf{ Improvement(\%)}\\ \hline
Method one\cite{r23,r24}&$0.1$&$2.7665$&$92.85$&$0.1610$&$0.3192$ &$0.0070$&$24.29$ \\
Method two\cite{r16,r38,r39}&$0.1$&$3.2193$&$80.71$&$0.1874$&$0.0000$&$0.0029$&$200$\\
Our work&$0.1$&$2.8165$&$91.34$&$0.1639$&$0.0872$&$0.0087$&-\\
\bottomrule
\end{tabular}
\end{table}

\begin{table}[!t]
\tabcolsep 25pt 
\caption{\label{tab:tab2}The degree distribution functions of three error correction matrices with different $R$.}
\begin{tabular}{lll}
\toprule
\bm{$R$}&\textbf{Degree distribution function}&{\textbf{Threshold}}\\ \hline
0.05&$v=0.04r_1x^2_1x_2^{34}+0.03r_1x_1^3x_2^{34}+0.93r_1x_3$&3.674\\
&$u=0.01x_1^8+0.01x_1^9+0.41x_2^2x_3+0.52x_2^3x_3$&\\

0.1&$v=0.0775r_1 x^2_1 x_2^{20}+0.0475r_1x_1^3x_2^{22}+0.875r_1x_3$&2.541\\
&$u=0.0025x_1^{11}+0.0225x_1^{12}+0.03x_2^2 x_3+0.845x_2^3 x_3$&\\

0.15&$v=0.0858r_1x^2_1x_2^{12}+0.0996r_1x_1^3x_2^{14}+0.8146r_1x_3$&2.038\\
&$u=0.0160x_1^{10}+0.0194x_1^{16}+0.0198x_2^2x_3+0.7948x_2^3x_3$&\\
\bottomrule
\end{tabular}
\end{table}

We further analyzed the fluctuation of $V_A^{opt}$ with different $\xi$, $v_{el}$, and $R$. The simulation results in Fig.~\ref{fig6}(a) show that $V_A^{opt}$ and the corresponding maximized SKR decrease as $\xi$ increases. Within the fluctuation range of $\xi$, the change in $V_A^{opt}$ is very small. The simulation results in Fig.~\ref{fig6}(b) show that the $V_A^{opt}$ increases as $v_{el}$ increases and the corresponding SKR decreases. Within the fluctuation range of $v_{el}$, the change in the SKR is also very small. The above results indicate that our method is robust even if the system parameters fluctuate slightly in real time.

\begin{figure}
    {\label{fig6a}{\includegraphics[width = 8cm]{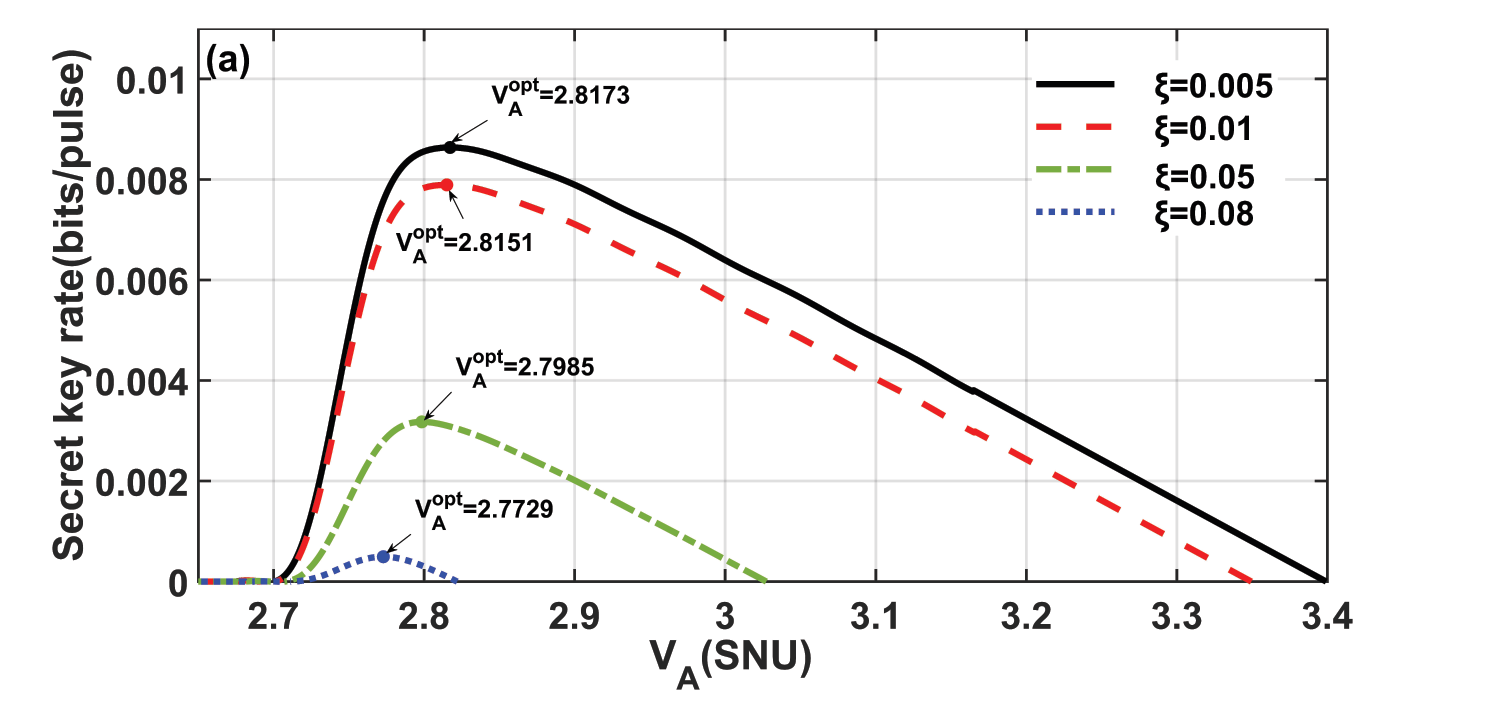}}}
    {\label{fig6b}{\includegraphics[width = 8cm]{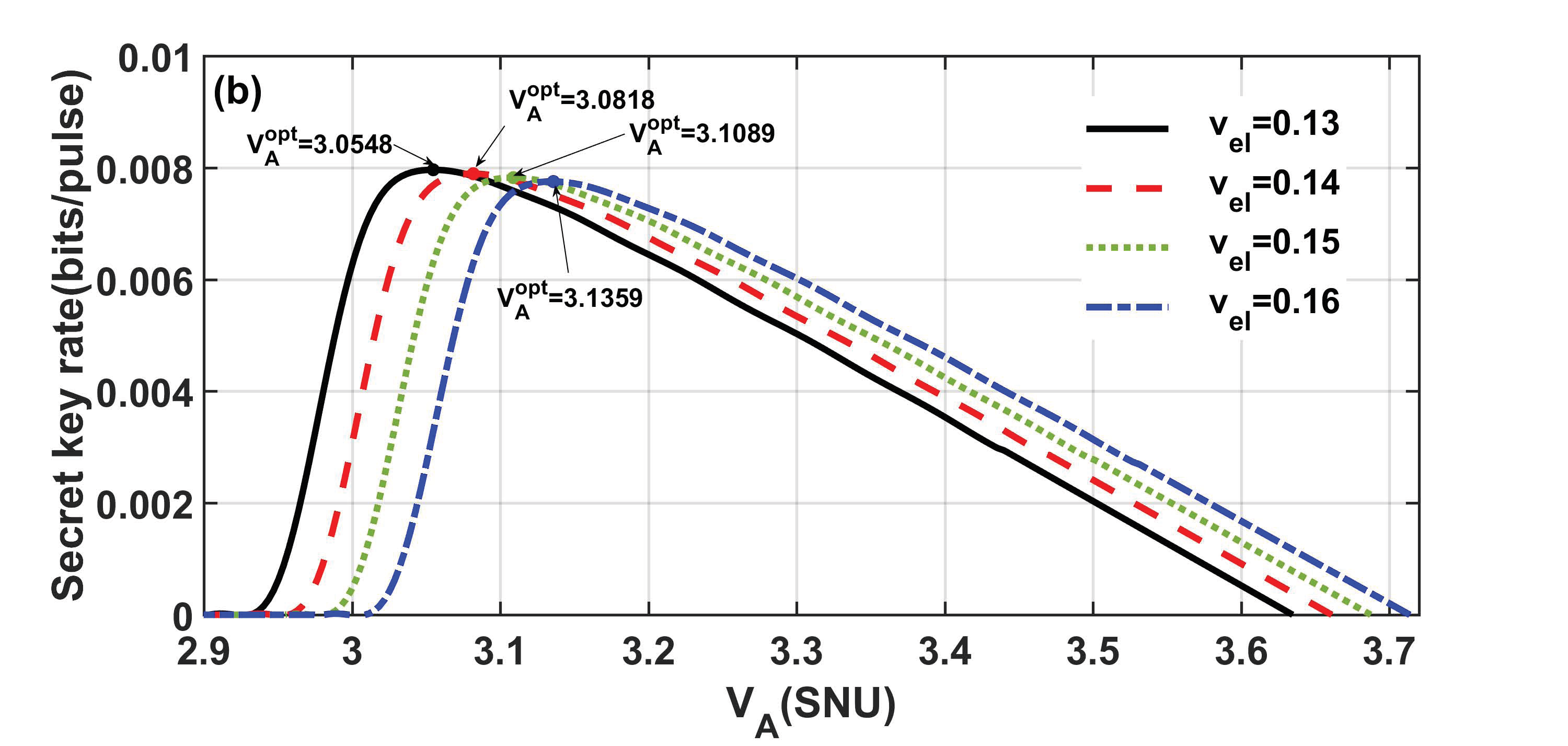}}}
    \caption{\label{fig6}Simulation results of the SKR with respect to $V_A$ optimization in homodyne detection: $\eta$=0.606, $\alpha$=0.2 $dB/km$, $R$=0.1, $L$=50 $km$. (a)$v_{el}$=0.041, $\xi$=0.005, 0.01, 0.05, 0.08 from top to bottom. The result of $V_A^{opt}$ is 2.8173, 2.8151, 2.7985, and 2.7729. (b) $\xi$=0.005, $v_{el}$=0.13, 0.14, 0.15, 0.16 from left to right. The result of $V_A^{opt}$ is 3.0548, 3.0818, 3.1089, and 3.1359. SNU: shot noise units.}{\label{fig6}}
\end{figure}

\begin{figure}
\centerline{
\includegraphics[width = 16cm]{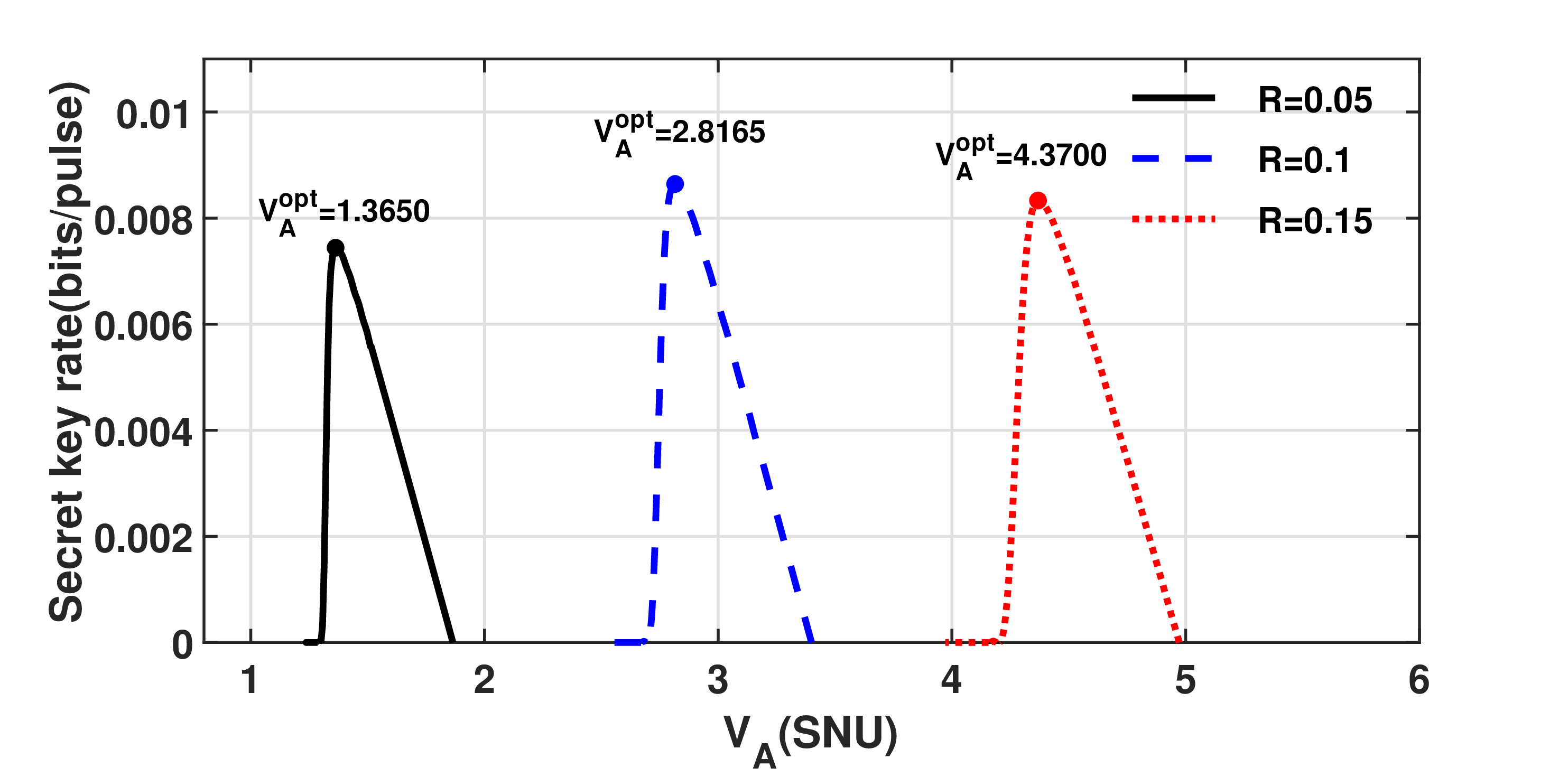}}
\caption{\label{fig7}SKR with respect to $V_A$ optimization under different code rate error corrections in homodyne detection: $v_{el}$=0.041, $\eta$=0.606, $\alpha$=0.2 $dB/km$, $L$=50 $km$, $\xi$=0.005, and $R$=0.05, 0.1, 0.15 from left to right. The results of $V_A^{opt}$ are 1.3650, 2.8165, and 4.3700. The result of the SKR is 0.0074 $bits/pulse$, 0.0086 $bits/pulse$, and 0.0083 $bits/pulse$, respectively. SNU: shot noise units.}
\end{figure}

The optimal choice of \textbf{H} is a complex problem. To verify the systematic optimization method between $V_A$ and \textbf{H} choice, the density evolution method \cite{r41} was chosen to generate the degree distribution function for $R$=0.05, 0.1, 0.15, as shown in Table ~\ref{tab:tab2}. Then, the progressive-edge-growth algorithm \cite{r42} was chosen to generate the matrix \textbf{H} according to the degree distribution functions, whose FERs were experimentally measured and fitted with a nonlinear function similar to the results in Eqs.~(\ref{eq6}) and ~(\ref{eq7}). The SKR of a CV-QKD system with different \textbf{H} at a transmission distance of 50 $km$ is shown in Fig.~\ref{fig7}, where the number of iterations is 60. Based on the above simulation results, to gain the maximum SKR, the matrices with $R$=0.1 are the best choice. However, for a practical CV-QKD system, the optimal $V_A$ influences not only the SKR but also the difficulty in quantum state preparation, measurement, and postprocessing. Accordingly, the proposed method in this paper provides a tool to calculate the cost quantitatively if the matrices with the maximum SKR are not chosen.

\begin{figure}
\includegraphics[width = 16cm]{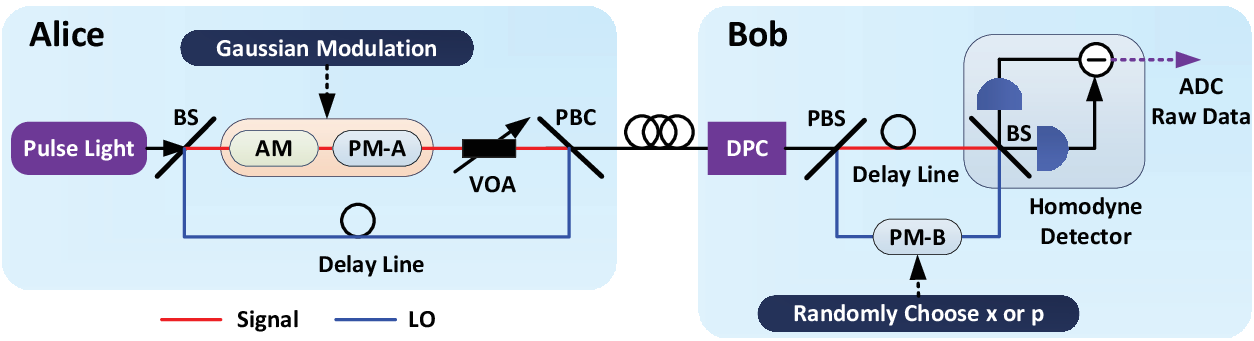}
\caption{\label{fig8}Sketch of the CV-QKD system based on the GG02 protocol. AM: amplitude modulator, PM-A/PM-B: phase modulator, DPC: dynamic polarization controller, VOA: variable optical attenuator, BS: beam splitter, PBS/C: polarization beam splitter/coupler, LO: local oscillator, ADC: analog-to-digital converter.}
\end{figure}

\begin{figure}
\includegraphics[width = 16cm]{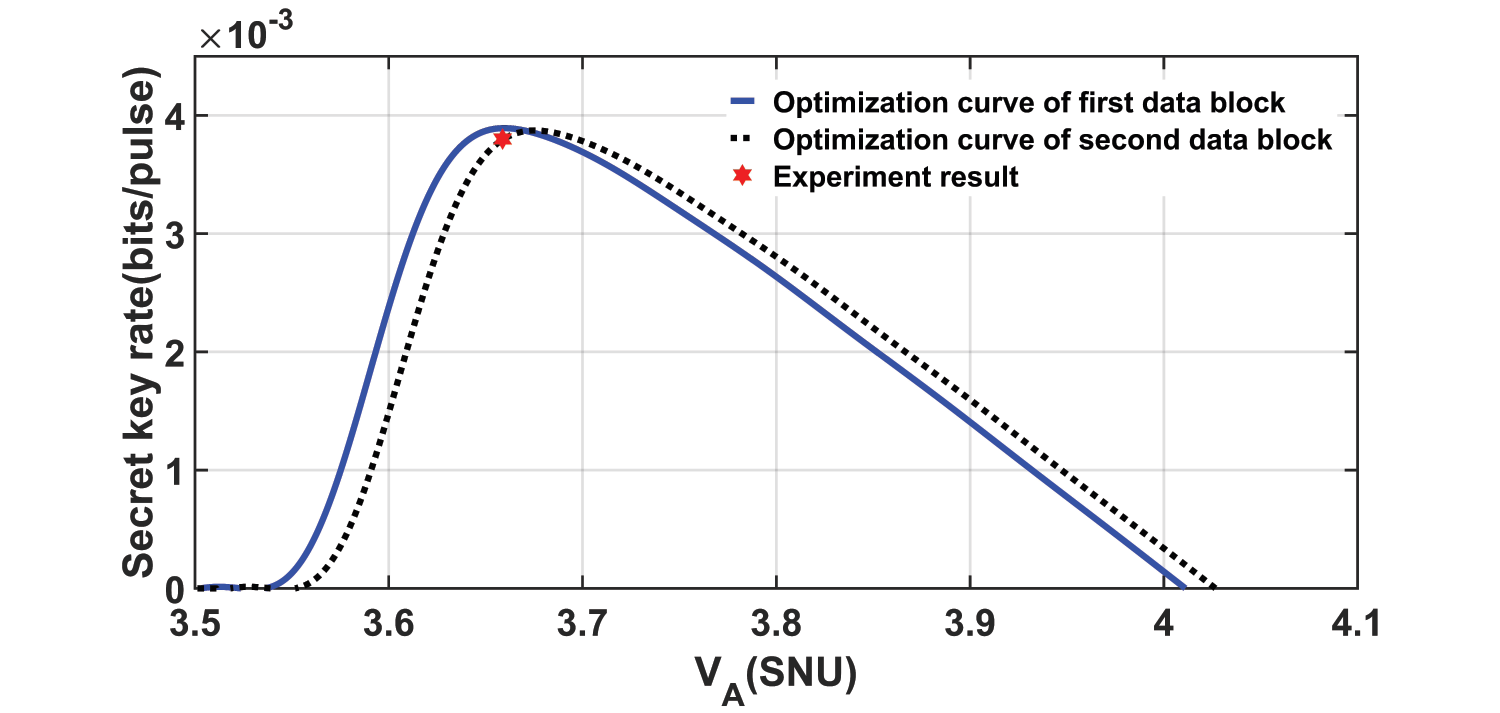}
\caption{\label{fig9}Experiment results of the SKR with respect to $V_A$ optimization. The parameters based on the first data block: $T$=0.1, $v_{el}$=0.1465, $\eta$=0.51, $R$=0.1, and $\xi$=0.0324. The parameters based on the second data block: $T$=0.1, $v_{el}$=0.1507, $\eta$=0.51, $R$=0.1, $V_A$=3.6588, $\xi$=0.0321, and $K_{experiment}^{opt}$=0.00380 $bits/pulse$ (red dot). The blue line is the optimized curve based on the first data block parameters, where $V_A^{opt}$=3.6608 and $K_{first}^{opt}$=0.00385 $bits/pulse$, and the dashed line is the optimized curve based on the second data block parameters for reference, where $V_A^{opt}$=3.6746 and $K_{second}^{opt}$=0.00386 $bits/pulse$.}
\end{figure}

\subsection{\label{sub:level4}Experimental verification}
We built a CV-QKD system with the GG02 protocol to verify the proposed methods experimentally, as shown in Fig.~\ref{fig8}. At Alice's side, a pulse light was separated into local oscillator (LO) light and signal light by an asymmetrical Mach-Zehnder interferometer (AMZI), where the $x$ and $p$ quadratures of the signal light are modulated with a Gaussian distribution by an amplitude modulator (AM) and phase modulator (PM-A). Based on polarization-multiplexing and time-multiplexing methods, the signal light was cotransmitted with LO light to Bob through a fiber channel. At Bob's side, the time and polarization de-multiplexing between LO and signal light was realized by a dynamic polarization controller (DPC), polarization beam splitter (PBS), and matched AMZI. Then, Bob randomly measured either $x$ or $p$ quadrature of signal light by a shot-noise-limited homodyne detector. The output signal of the homodyne detector, which is proportional to the modulation value of the quadrature for signal light, was acquired by an analog-to-digital converter (ADC) to obtain the raw data.

In our experiment, the transmission distance was 50 $km$, the block size $N$ was set as $128\times10^6$, and $n$ was set as $N/2$. The initial system parameters were estimated from the first data block, where $T$=0.1, $v_{el}$=0.1465, $\eta$=0.51, $R$=0.1, and $\xi$=0.0324, which simulated the first launch of the CV-QKD system in a practical optical fiber condition. Based on the proposed method, the optimal value can be calculated, where $V_A^{opt}$=3.6608. Then, we adjusted the modulation variance to $V_A^{opt}$. However, an inevitable slight parameter fluctuation could occur after the modification of the modulation variance, and we also cannot control $V_A$ with arbitrary accuracy in the experiment, where the system parameters in the second block turn to $T$=0.1, $v_{el}$=0.1507, $\eta$=0.51, $R$=0.1, $V_A$=3.6588, and $\xi$=0.0321. The obtained SKR for the second data was $K_{experiment}^{opt}$=0.00380 $bits/pulse$ (red dot). As shown in Fig.~\ref{fig9}, the blue line is the optimized curve based on the first data block parameters, where $V_A^{opt}$=3.6608 and the ideal theoretical optimal SKR $K_{first}^{opt}$=0.00385 $bits/pulse$. The dashed line is the optimized curve based on the second data block parameters, where $V_A^{opt}$=3.6746 and the ideal theoretical optimal SKR $K_{second}^{opt}$=0.00386 $bits/pulse$. By comparing $K_{experiment}^{opt}$, $K_{first}^{opt}$, and $K_{second}^{opt}$, the deviation of the experimentally obtained SKR is $<$1.6\%, which shows that the proposed method is feasible and robust for a practical system even with parameter fluctuations, and the available optimized result is very close to the ideal theoretical result.

\section{\label{sec:level4}Conclusion}
In conclusion, we propose a systematic optimization method for a practical CV-QKD system with a restricted capacity of postprocessing, and the feasibility was verified theoretically and experimentally. Our simulation results show that the SKR can be improved by 24\% and 200\% with the proposed method compared with previous frequently used optimization methods with a transmission distance of 50 $km$. The experimental results demonstrate that the method is feasible and robust to be applied in an actual CV-QKD system, where the deviation between the experimentally obtained SKR and the ideal optimal value is $<$1.6\% under system parameter fluctuation. Furthermore, the selection of optimal error correction matrices was studied with the proposed method, which provides a quantitative method to calculate the cost of the SKR if suboptimal matrices are chosen to reduce the decoding complexity in a practical CV-QKD system. This paper presents a method to improve the performance of CV-QKD systems in the field without modification of the hardware, which paves the way to deploy high-performance CV-QKD in the real world. Our method can also be effectively combined with other theoretical optimization methods, such as rate-adaptive algorithms, postselection, and add noise methods, which can be studied in the future.

\Acknowledgements{This work was supported in part by the National Key Research and Development Program of China (Grant No. 2020YFA0309704), the National Natural Science Foundation of China (Grant Nos 61901425, U19A2076, 62101516, 62171418, 62201530), the Sichuan Science and Technology Program (Grant Nos 2022ZYD0118, 2023JDRC0017, 2023YFG0143, 2022YFG0330, 2022ZDZX0009 and 2021YJ0313), the Natural Science Foundation of Sichuan Province (Grant Nos 2023NSFSC1387 and 2023NSFSC0449), the Basic Research Program of China(Grant No. JCKY2021210B059), the Equipment Advance Research Field Foundation(Grant No. 315067206), the Chengdu Major Science and Technology Innovation Program (Grant No. 2021-YF08-00040-GX), the Chengdu Key Research and Development Support Program (Grant Nos 2021-YF05-02430-GX and 2021-YF09-00116-GX), the Foundation of Science and Technology on Communication Security Laboratory (Grant No. 61421030402012111).
}

\Supplements{Appendix A.}


\begin{appendix}
\section{\label{app:subsec}Secret key rate calculation}

In Eq.~(\ref{eq1}) and Eq.~(\ref{eq3}), $I^{hom/het}(x:y)$ can be calculated as\cite{r36,ra7}

\begin{equation}\label{app1}
f_{I^{hom/het}(x:y)}(V_A)=I^{hom/het}(x:y)=\frac{v_{det}}{2}\log_{2}\left(\frac{V+\chi_{tot}^f}{1+\chi_{tot}^f}\right)
\end{equation}
where $V=V_A+1$, and $\chi_{tot}^f$ represents the total noise referred to the channel input, $\chi_{tot}^f=\chi_{line}^f+\chi_{hom/het}/T_{min}$, and $\chi_{line}^f=(1-T_{min})/T_{min}+\xi_{max}$ is the total channel added noise referred to the channel input, and $\chi_{hom/het}=v_{det}(1+v_{el})/\eta-1$ is the total added noise introduced by the realistic homodyne/heterodyne detector referred to Bob’s input. It is proved in Refs.\cite{r36,ra7} that

\begin{equation}\label{app2}
T_{min}=\frac{\tau_{min}}{\eta}=\frac{\tau-\Delta\tau}{\eta}=\frac{\tau}{\eta}-\frac{2w}{\eta}\sqrt{\frac{2\tau^2V_A+\tau\sigma^2_z}{m_pV_A}}
\end{equation}
\begin{equation}\label{app3}
\xi_{max}=\xi+\frac{2w\sigma^2_z}{\tau\sqrt{2m_p}}
\end{equation}
where $\tau=\eta T$ and $\sigma_z^2=\eta T\xi+v_{det}v_{el}+v_{det}$. Alice and Bob randomly and jointly choose $m$ of the $N$ distributed signals for parameter estimation, and the corresponding $m_p=v_{det}m$. $v_{det}$ is the quantum duty ("qu-duty") associated with detection: $v_{det}=1$ for homodyne and $v_{det}=2$ for heterodyne. Confidence parameter $w$ is determined by the tolerable error probability $\varepsilon_{pe}$, which typically set as $w$=6.34, $\varepsilon_{pe}=2^{-33}$\cite{r36,ra7}.

$\chi^{hom/het}(y:E)$ can be estimated as \cite{r36,ra7}

\begin{equation}\label{app4}
f_{\chi^{hom/het}(y:E)}(V_A)=\chi^{hom/het}(y:E)=\sum_{i=1}^{2}G\left(\frac{\lambda_i-1}{2}\right)- \sum_{i=3}^{5}G\left(\frac{\lambda_i-1}{2}\right)
\end{equation}
where $G(x)=(x+1)\log_2(x+1)-x\log_2x$, and $\lambda_i$ are the symplectic eigenvalues of the covariance matrix $\gamma_{AB}$ between Alice and Bob. $\lambda_{1,2}$ is given by

\begin{equation}\label{app5}
\lambda_{1,2}^2=\frac{1}{2}[A\pm\sqrt{A^2-4B}]
\end{equation}
Similarly, the $\lambda_{3,4,5}$ is given by

\begin{equation}\label{app6}
\lambda_{3,4}^2=\frac{1}{2}[C\pm\sqrt{C^2-4D}], \lambda_5=1
\end{equation}
In GG02 protocol with homodyne detection,
\begin{equation}\label{app7}
\begin{aligned}
A=&V^2(1-2T_{\min})+2T_{\min}+ T_{\min} ^2(V+\chi_{line}^f)^2\\
B=&T_{\min}^2(V\chi_{line}^f+1)^2\\
C=&\frac{V\sqrt{B}+T_{\min}(V+\chi_{line}^f)+A\chi_{hom}}{T_{min}(V+\chi_{tot}^f)}\\
D=&\frac{\sqrt{B}(V+\sqrt{B}\chi_{hom})}{T_{min} (V+\chi _{tot}^f)}\\
\end{aligned}
\end{equation}
and in the no-switching protocol with heterodyne detection,
\begin{equation}\label{app8}
\begin{aligned}
A=&V^2(1-2T_{\min})+2T_{\min}+T_{\min}^2{(V+\chi_{line}^f)}^2\\
B=&T_{\min}^2{ (V\chi _{line}^f+1)}^2 \\ 
C=&\frac{1}{(T_{\min}(V+\chi_{tot}^f))^2}[A\chi_{het}^2+B+1+2\chi_{het}(V\sqrt{B}+T_{\min}(V+\chi_{line}^f))+2T_{\min} (V^2-1)] \\
D=&{\left(\frac{V+\sqrt{B}\chi_{het}}{T_{\min} (V+\chi _{tot}^f)}\right)}^2 \\ 
\end{aligned}
\end{equation}

\end{appendix}

\end{document}